\documentclass[preprints,article,accept,pdftex,moreauthors]{Definitions/mdpi} 
\firstpage{1} 
\pubvolume{8}
\issuenum{11}
\articlenumber{570}
\pubyear{2022}
\copyrightyear{2022}
\externaleditor{Academic Editor: Maciej \linebreak Rybczynski} 
\datereceived{19 September 2022} 
\dateaccepted{21 October 2022} 
\datepublished{30 October 2022} 
\hreflink{https://doi.org/10.3390/\linebreak universe8110570} 
\def\nat{Nature\ }
\def\aap{Astron.\ Astrophys.\ }
\def\apj{Astrophys.\ J.\ }
\def\apjl{Astrophys.\ J.\ Lett.\ }

\def\physrep{Phys.\ Rept.\ }
\def\prd{Phys.\ Rev.\ D\ }
\def\prl{Phys.\ Rev.\ Lett.\ }

\def\araa{Annu.\ Rev.\ Astron.\ Astrophys.\ }

\pdfoutput=1

\Title{An Unsupervised Machine Learning Method for Electron--Proton 
 Discrimination of the DAMPE Experiment}

\TitleCitation{\mbox{An Unsupervised} Machine Learning Method for Electron--Proton Discrimination of the DAMPE Experiment}

\Author{{Zhihui Xu $^{1,2}$\orcidC{},} 
 Xiang Li $^{1,2,}$*\orcidA{}, Mingyang Cui $^{1,}$*, {Chuan Yue $^{1}$, Wei Jiang $^{1}$, Wenhao Li $^{1,2}$} and Qiang Yuan $^{1,2}$\orcidB{}}

\AuthorNames{Zhihui Xu, Xiang Li, Mingyang Cui, Chuan Yue, Wei Jiang, Wenhao Li and Qiang Yuan}

\AuthorCitation{Xu, Z.; Li, X.; Cui, M.; Yue, C.; Jiang, W.; Li, W.; Yuan, Q.}

\address{%
$^{1}$ \quad Key Laboratory of Dark Matter and Space Astronomy, Purple Mountain Observatory, Chinese Academy of Sciences, Nanjing 210023, China 
\\
$^{2}$ \quad School of Astronomy and Space Science, University of Science and Technology of China, Hefei 230026, China}

\corres{Correspondence: xiangli@pmo.ac.cn (X.L.); mycui@pmo.ac.cn (M.C.)}

\abstract{Galactic cosmic rays are mostly made up of energetic nuclei, with less than $1\%$ of electrons
(and positrons). Precise measurement of the electron and positron component requires a very efficient
method to reject the nuclei background, mainly protons. In this work, we develop an unsupervised 
machine learning method to identify electrons and positrons from cosmic ray protons for the Dark 
Matter Particle Explorer (DAMPE) experiment.
Compared with the supervised learning method used in the DAMPE experiment, this unsupervised method relies solely on real data except for the background estimation process. As a result, it could effectively reduce the uncertainties from simulations.
For three energy ranges of electrons and positrons,  80--128 GeV, 350--700 GeV, and \mbox{2--5 TeV}, the
residual background fractions in the electron sample are found to be about (\mbox{0.45 $\pm$ 0.02)$\%$},
(0.52 $\pm$ 0.04)$\%$, and (10.55 $\pm$ 1.80)$\%$, and the background rejection power is about
(6.21 $\pm$ 0.03) $\times$ $10^4$, (9.03 $\pm$ 0.05) $\times$ $10^4$, and (3.06 $\pm$ 0.32) $\times$ $10^4$, respectively. 
This method gives a higher background rejection power in all energy ranges than the traditional 
morphological parameterization method and reaches comparable background rejection performance 
compared with supervised machine learning~methods.}

\keyword{\textls[-25]{DAMPE; machine learning; principal component analysis; particle identification; \mbox{cosmic rays}}} 

\begin{document}


\section{Introduction}
\label{intro}

Electrons\endnote{\scalebox{.9}[1.0]{(Throughout this paper, we use electrons to represent electrons and positrons
without discriminating them unless specified~otherwise.)}}  
 in cosmic rays (CR) are important probe of nearby CR accelerators due to their short propagation distances in the Milky Way
\cite{1995PhRvD..52.3265A,2018SCPMA..61j1002Y}. 
They are also widely used to search for new physics, such as the particle dark matter
\cite{2010ARA&A..48..495F, 2005PhR...405..279B}. The abundance of CR electrons above GeV is significantly lower, by a factor of $10^{-3}$$\sim$$10^{-2}$, than that of CR protons. 
Therefore, it is challenging to precisely measure the spectrum of electrons. Currently, the best 
measurements of the electron and/or positron spectra come from space (or balloon) direct 
detection experiments, including the magnetic spectrometers and imaging calorimeters
\cite{2001ApJ...559..296D,2001ApJ...559..973T,2008Natur.456..362C,2014PhRvL.113v1102A,2016ApJ...831...18C,2017PhRvD..95h2007A,2017Natur.552...63D,2017PhRvL.119r1101A}. 
The ground-based atmospheric imaging Cherenkov telescope arrays also tried to measure the total
electron plus positron spectra to higher energies, which, however, are subject to large systematic
uncertainties \cite{2008PhRvL.101z1104A,2009A&A...508..561A,2011ICRC....6...47B,2015ICRC...34..411S}.
The spectra of electrons were measured up to a few TeV, experiencing a softening around a few
GeV, a hardening around $50$ GeV, and a softening around $0.9$ TeV \cite{2018SCPMA..61j1002Y}.
Those spectral features help establish a three-component origin model of electrons and positrons, 
including primary electrons from CR acceleration sources, secondary electrons and positrons
from inelastic interactions between CR nuclei and the interstellar medium, and additional
electrons and positrons relevant to the high-energy excesses \cite{2018SCPMA..61j1002Y}.

The DArk Matter Particle Explorer (DAMPE) is a space high-energy charged CR and gamma-ray detector
optimized for precision detection of electrons with a very high energy resolution and background
rejection \cite{ChangJin:550,2017APh....95....6C}. DAMPE is a calorimetric-type instrument, which 
consists of four sub-detectors. 
The Plastic Scintillator Detector (PSD; \cite{2017APh....94....1Y}) on the top is used to measure 
the particle charge up to $Z=28$, and serves as an anti-coincidence detector for $\gamma$ rays.
The charge resolution of PSD was found to be about $0.137$ (full width at half maximum) for $Z=1$
\cite{2019APh...105...31D}. 
The Silicon Tungsten tracKer-converter (STK) is designed for the trajectory measurement
\cite{2016NIMPA.831..378A}. It can also measure the particle charge for $Z<8$. 
The {${\rm Bi}_4{\rm Ge}_3{\rm O}_{12}$} (BGO; \cite{2015NIMPA.780...21Z}) calorimeter plays a 
crucial role in the energy measurement and the electron--proton discrimination. The total thickness 
of the BGO calorimeter of DAMPE reaches $\sim$32 radiation lengths and thus enables the calorimeter to contain electromagnetic showers without remarkable leakage below $\sim$10 TeV, which ensures a very high energy resolution 
(better than $1.5\%$ for $E>10$ GeV) and a high electron--proton discrimination capability. 
The NeUtron Detector (NUD; \cite{2016AcASn..57....1H}) at the bottom provides a further 
electron--proton separation via the detection of secondary neutrons produced by interactions in 
the calorimeter. All the detectors have operated stably in space since the launch of DAMPE in December,
2015 \cite{2019NIMPA.924..309T,2019APh...106...18A}. Important progress in measuring the
electron and CR nuclear spectra has been achieved \cite{2017Natur.552...63D,2019SciA....5.3793A,2021PhRvL.126t1102A,2021ApJ...920L..43A,2021arXiv211208860A}.

One of the most important elements for precise measurements of the electron spectrum is to
``suppress'' the proton background. For a calorimeter detector, this can be achieved by means of 
the shower morphology differences between hadronic showers and electromagnetic showers.
Typically, an electromagnetic shower spreads less with a more regular morphology than a 
hadronic shower with similar deposited energy. In Ref.~\cite{2017Natur.552...63D}, a
two-parameter representation of the shower morphology was developed, i.e., the lateral 
spread and the longitudinal development. This method can effectively suppress the proton
background\endnote{(note that heavier nuclei can be highly suppressed by the charge
measurement, leaving protons as the main background)}, resulting in the level
of background for electron energies a  few percent below TeV. However, for $E>$ TeV, the background
increases quickly for this traditional method. An optimization of the electron--proton
discrimination is necessary (e.g., \cite{2021JInst..16P7036D,2018RAA....18...71Z}).

The Principal Component Analysis (PCA) is one of the most commonly used machine learning 
methods for dimensionality reduction and feature extraction \cite{1999ASPC..162..363F,10.1093/biomet/87.3.587,6790375, 2008ITSP...56.5823U}. 
The working principle of PCA is to express the original data in a new data space.
 Compared to other machine learning methods, PCA is an unsupervised machine learning and 
 thus does not rely on simulated data. Therefore, it may avoid potential biases from models 
 of simulation. The disadvantage of the PCA method is that the limited data statistics at
 high energies may result in relatively large statistical uncertainties.

 This work develops an algorithm to separate electrons from protons using the PCA method. 
In Section~\ref{Sec_PCA_method}, we introduce the basic principle of the PCA method. 
In Section~\ref{Sec_Electron_identification}, we present the detailed algorithm to separate 
electrons from protons applicable to the DAMPE experiment. Section~\ref{Sec_Performance} gives 
the results of our method. Finally, we conclude this work in Section~\ref{Sec_Conclusion}.
 
\section{The PCA Method}
\label{Sec_PCA_method}
Generally speaking, the PCA method corresponds to a transform in a high-dimensional (not necessarily
orthogonal) parameter space through a rotation matrix to find a new coordinate system, in which
the variances of the data along the major axes of the new coordinate are the largest. Larger variance 
means that the data are more discrete and discriminative. Finding the coordinate axes corresponding to the maximum variance is equivalent to determining the eigenvectors corresponding to the maximum eigenvalues of the covariance matrix of the original data. 
The commonly used method to 
solve the eigenvalues of the covariance matrix is the Singular Value Decomposition method \cite{https://doi.org/10.1137/090771806,MARTINSSON201147,Minka2000AutomaticCO}.


In our analysis, we characterize the shower morphology as the energy deposition ratio and the 
hit dispersion (see Section \ref{Sec_Electron_identification} for more details) in each BGO layer. 
A vector space is formed by the linear combination of these variables, which are then transformed 
into a new space through a linear transformation. 
In the new vector space, the first several principal components retain most of the variance of the data sample.
In this work, we keep only the first three components and ignore the others.
In summary, our analysis consists of 5 steps:
\begin{enumerate}[label=(\arabic*)]
    \item Selecting the data with good reconstruction.
    \item Constructing characteristic variables carrying shower morphology information.
    \item Finding the eigenvector and transformation matrix.
    \item Transforming the original data into the new space and finding the first three principal~components.
    \item Rotating the previous three-dimensional space to obtain the final component to discriminate 
    electrons from protons.
\end{enumerate}



\section{Electron-Proton Separation}
\label{Sec_Electron_identification}

\subsection{Data Selection}
Six years of  DAMPE flight data are used in this analysis. The instrument dead time after trigger, 
the on-orbit calibration time, and the time when the satellite passes through the South Atlantic Anomaly
region are excluded. We first apply a pre-selection procedure to select events with an accurate track
reconstruction and a good shower containment in the BGO calorimeter. This procedure consists of a few
specific conditions as follows:
\begin{itemize}
    \item The events should meet the High Energy Trigger (HET) {\cite{2017Natur.552...63D}} condition to ensure a good shower development at the beginning of the BGO caloriment.
    \item The radial spread of the shower development, defined as the Root Mean Square (RMS) of the distances between the hit BGO bars and the shower axis, \linebreak{}\mbox{${\rm RMS}_r=\sqrt{\sum_{j=1}^{N}E_{j}\times D_j^2 /E_{\rm total}}$,} should be smaller than 40 mm. 
    The $E_j$ is energy deposited in $j$-th BGO bar, and $D_j$ is the distance between the corresponding BGO bar and track of the particle.
    This cut could eliminate a large fraction of nuclei because the hadronic shower is typically wider than the electromagnetic one.
    \item The max energy bar of the BGO should not be on the edge of the detector. 
    \item The max energy ratio of each layer, e.g., the ratio of the max energy of a single BGO bar over the total energy of that layer, should be less than 0.35.  The cut can eliminate those particles coming from the side of the detector.
    \item The reconstructed track should pass through the top and bottom surfaces of the BGO.
    \item Events with PSD charge should be smaller than 2 to remove heavy nuclei.
\end{itemize}

\begin{figure}[H]
\includegraphics[width=9 cm]{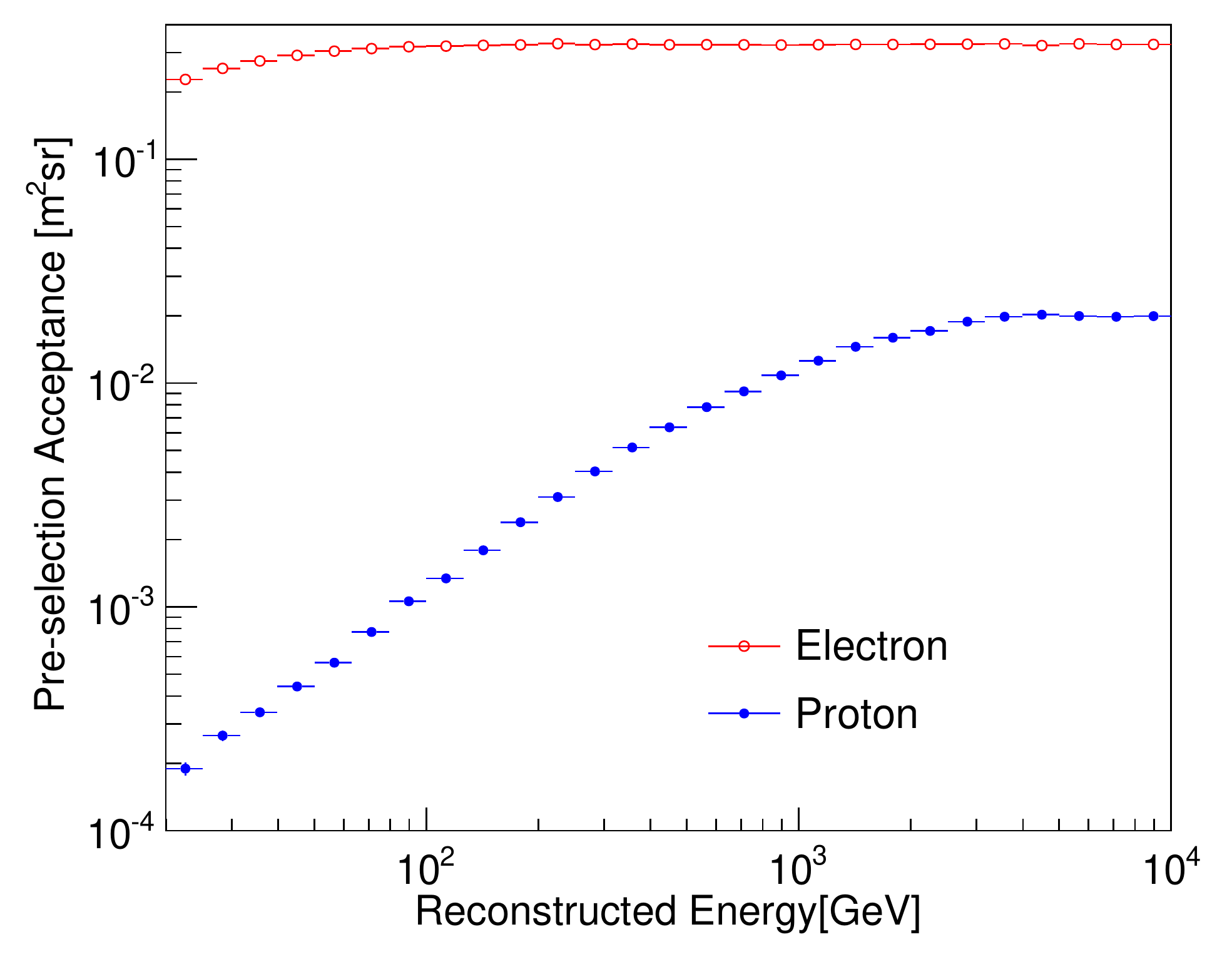}
\caption{{The pre-selection} 
 acceptance of electrons and protons.\label{fig:Pre_eff}}
\end{figure}

We show the efficiency of these pre-selection conditions in Figure~\ref{fig:Pre_eff}. 
The results are obtained from the Monte Carlo (MC) simulation for an isotropic source
distribution with 1~m radius and $E^{-1}$ spectrum. The spectrum is then
re-weighted to $E^{-2.7}$ for protons and $E^{-3.1}$ for electrons.
We see that the pre-selection procedure can be able to suppress protons by a factor of $10$$\sim$$10^3$, 
mainly due to the HET requirement. Furthermore, since the CR proton spectrum is approximately 
proportional to $E^{-2.7}$, the different energy deposition fractions in the calorimeter of 
protons ($30$$\sim$$50\%$) and electrons ($>$90$\%$) would contribute to the suppression factor 
by about 3$\sim$7 for a given reconstructed energy window {\cite{Yue_2017}}.
    \vspace{-6pt}
\unskip

\subsection{Construction of Characteristic Variables}
The BGO calorimeter is composed of 14 layers, and each layer consists of 22 BGO crystals placed in a hodoscopic configuration \cite{2015NIMPA.780...21Z}. With the hit information from those 308 BGO crystals, we characterize the shower morphology from longitudinal and lateral views, respectively.
The longitudinal shower development is characterized by the energy ratio in each BGO layer, 
$F_i=E_{i}/E_{\rm total}$, where $E_{i}$ is the deposited energy of the $i$-th layer and 
$E_{\rm total}$ is the total deposited energy in the calorimeter. 
The lateral spread, on the other hand, is described by the RMS of the energy deposits in each layer,
\begin{equation}
    {\rm RMS}_i=\sqrt{ \frac{ \sum_{j=1}^{22}E_{ij} \times (d_{ij}-d_{i}^{\rm cog})^{2}}{\sum_{j=1}^{22}{E_{ij}} }},\ i=0,\ \ldots,\ 13,   \label{RMS}
\end{equation}
where $E_{ij}$ is the deposited energy of the $j$-th bar in the $i$-th layer, $d_{ij}-d_i^{\rm cog}$ 
is the distance from the $j$-th bar in the $i$-th layer to the ``center of gravity'' of the 
$i$-th layer, defined as 
\begin{equation}
    d_{i}^{\rm cog}=\sum_{j=1}^{22}E_{ij} \times \frac{d_{ij}}{E_{i}}.  \label{cog}
\end{equation}


Based on these 28 basic variables, $F_i$ and RMS$_i$, we further construct higher-order variables 
to achieve a better particle discrimination.
The simplest way is to randomly weight RMS$_i$ and $F_i$ to form a new set of variables and to
search for optimal weighting coefficients. We define the new variables as
\begin{eqnarray}
    {\rm RMS}_i'&=&{\rm RMS}_i\times ({\rm cos} ~\theta)^{\gamma} \times \alpha_{i} \nonumber\\
    F_{i}'&=& F_i \times \beta_{i},
\end{eqnarray}
where $\theta$ is the angle between the reconstructed incident direction and the vertical 
direction of an event, and $\alpha_{i}$, $\beta_{i}$, $\gamma$ are random numbers between 0 and 1, 
which will be determined by the PCA.


\subsection{Finding the Principal Components}

The major task of the PCA analysis is to find the optimal weighting coefficients of the variables, 
i.e., $\alpha_{i}$, $\beta_{i}$ and $\gamma$. We first generate tens of millions of random sets 
of weighting parameters. For a set of random weights, there is a new vector $\{{\rm RMS}_i',\ F_i'\}$ for an event.
Then, a covariance matrix can be obtained for a data sample. 
The direction of the first principal component is the direction of the eigenvector corresponding 
to the largest eigenvalue of the covariance matrix. Mathematically, this is to solve the eigenvectors 
and eigenvalues of the covariance matrix. The eigenvectors, placed in descending order of 
eigenvalues, form the transformation matrix. Multiplied by this transformation matrix, the
vector $\{{\rm RMS}_i',\ F_i'\}$ is transformed to a new one $\{X,\,Y,\,Z,\,\ldots\}$, which gives
the principal components in descending order of their capabilities to distinguish particles. 
We find the transformation matrix using the python package 
{\tt sci-kit} {(\url{https://scikit-learn.org/}}{, accessed on 20 September 2022)} 
 and calculate the proton rejection power. 
The optimal condition is to ensure that the ratio between the peak of the distribution of 
electron candidates and the valley is as large \mbox{as possible.}

\begin{figure}[H]
\includegraphics[width=10.5 cm]{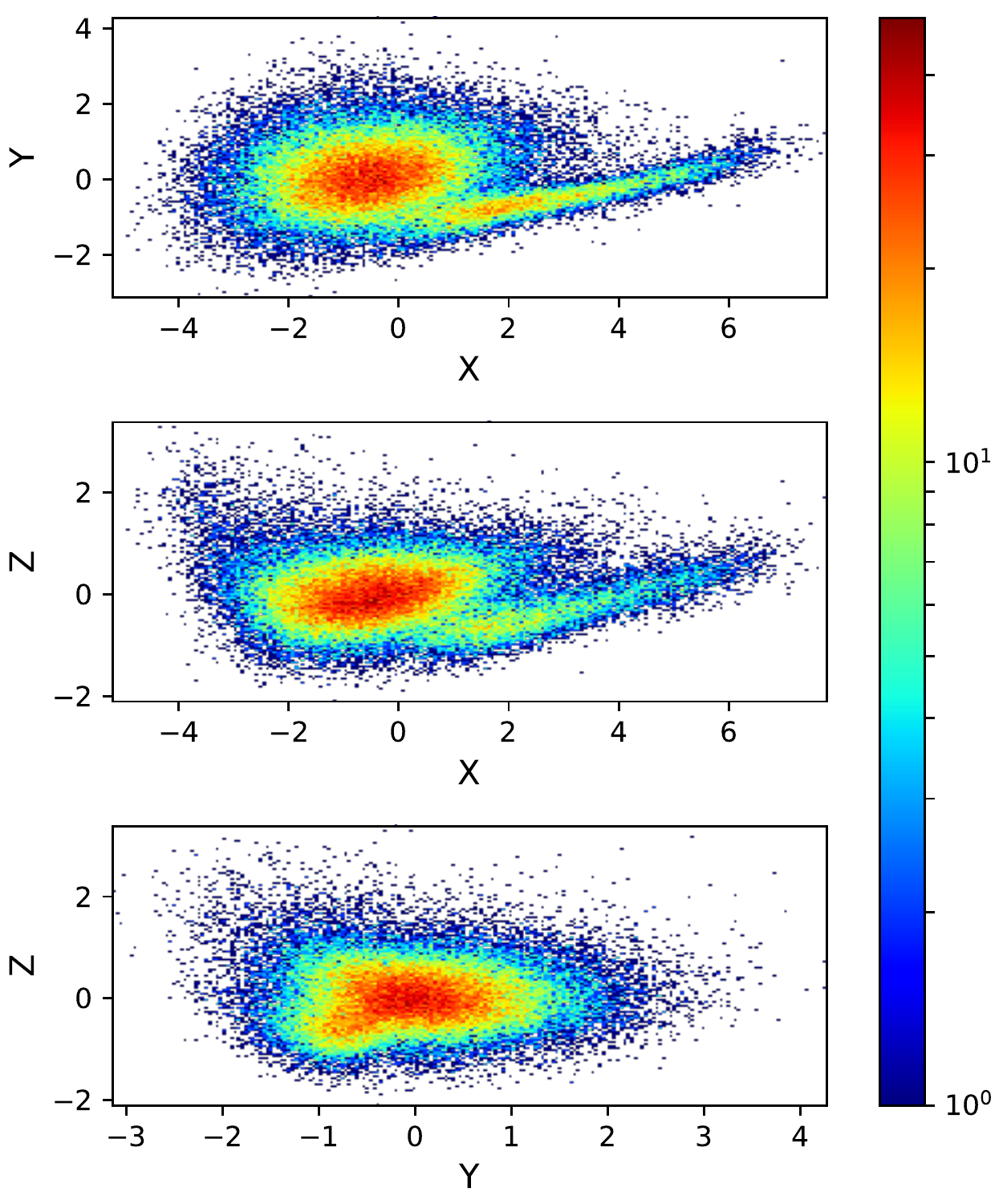}
\caption{The scattering plots of the first three principal components in the 350.0--700.0 
    GeV reconstructed energy range.\label{fig:3D_PCA}
}
\end{figure}

    
The output of the PCA is a vector group with an orthogonal rank reduction. The first principal 
component with the largest variance, however, may not be able to effectively distinguish electrons from protons by itself. 
We therefore keep the first three principal components. 
For simplicity, we choose the energy range of  350.0--700.0 GeV for illustration in this section. 
The scattering plots of the first three most
informative dimensions of the PCA components for reconstructed energies of 350.0--700.0 GeV are shown 
in Figure~\ref{fig:3D_PCA}. We use $X$, $Y$, and $Z$ to illustrate the first, second, and third principal
components. It shows that the $X$ component gives the relative better discrimination power of the electrons and protons.
For the $Z$ component, the two groups of events are almost indistinguishable. 
The corresponding parameters found for reconstructed energy of 350.0--700.0 GeV are shown in Table~\ref{param_tab}.

\vspace{-3pt} {}

\vspace{-6pt}
\begin{table}[H] 
\caption{The parameters found in the energy range of 350  to 700 GeV by the PCA method.\label{param_tab}}
\newcolumntype{C}{>{\centering\arraybackslash}X}
\begin{tabularx}{\textwidth}{CCC}
\toprule
\boldmath{$\alpha$}	& \boldmath{$\beta$}	& \boldmath{$\gamma$} \\
\midrule
0.3539   &    0.4676     &              \\
0.9451   &    1.535      &              \\
0.9551   &    1.723      &              \\
1.2974   &    0.2088     &              \\
0.06981  &    0.06027    &              \\
1.054    &    0.7731     &              \\
1.946    &    0.5759     &              \\
0.8407   &    0.07682    &      0.3755  \\
1.280    &    1.109      &              \\
1.414    &    1.695      &              \\
1.509    &    0.2808     &              \\
1.987    &    0.4533     &              \\
1.890    &    1.241      &              \\
0.7533   &    1.745      &              \\
\bottomrule
\end{tabularx}
\end{table}



\begin{figure}[H]
\includegraphics[width=10.5 cm]{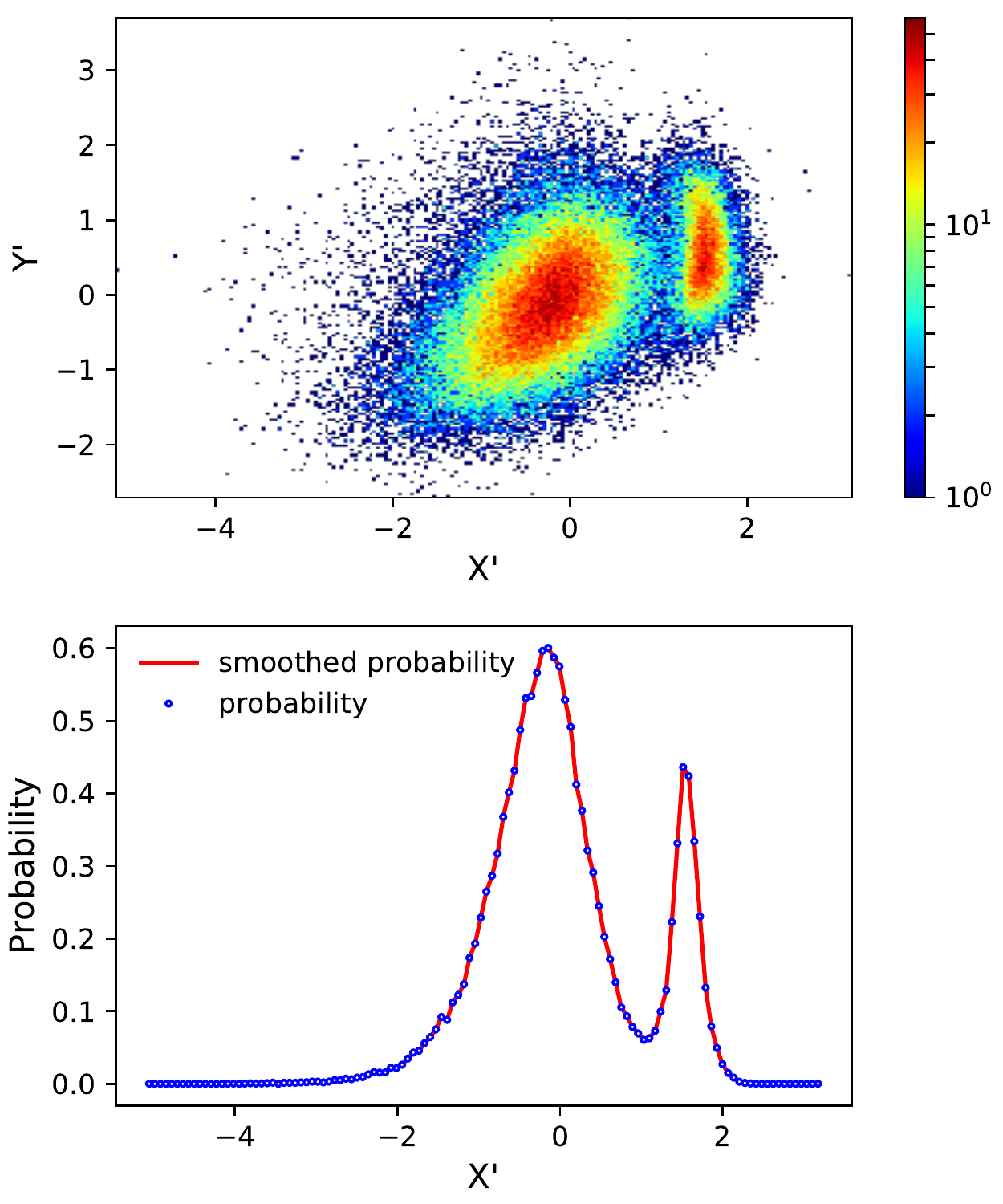}

\caption{The distribution of the $X^{\prime},Y^{\prime}$ in the 350.0--700.0 GeV reconstructed energy range.\label{fig:Pro_PCADis}}
\end{figure}

For the convenience of use of the PCA results, we further rotate the vector space of the first
three components to find a new principal direction, which distinguishes electrons from protons
most effectively. This is equivalent to seeking a rotation from ($X,Y,Z$)
to a new set of basis ($X^{\prime},Y^{\prime},Z^{\prime}$), such that the single $X^{\prime}$ is enough to discriminate electrons from protons well.
After a proper rotation, we obtain a clearer 
separation of electrons and protons using the new variable $X^{\prime}$, as shown in Figure~\ref{fig:Pro_PCADis}.


\unskip

\section{Results}\label{Sec_Performance}
    

Using the PCA method, we reduce the 28-dimensional parameter space to three major principal
components to form a new vector space. The three-dimensional vector space is then further
rotated to form a new principal axis, which separates the electrons from protons most effectively.
In order to estimate the performance of the electron--proton discrimination, we use the MC 
simulation samples of electrons and protons as templates to fit the flight data. Note that the transformation matrix is obtained directly from the flight data,
which makes our method distinct from the supervised machine learning.

\begin{figure}[H]
	\includegraphics[width=0.35\textwidth]{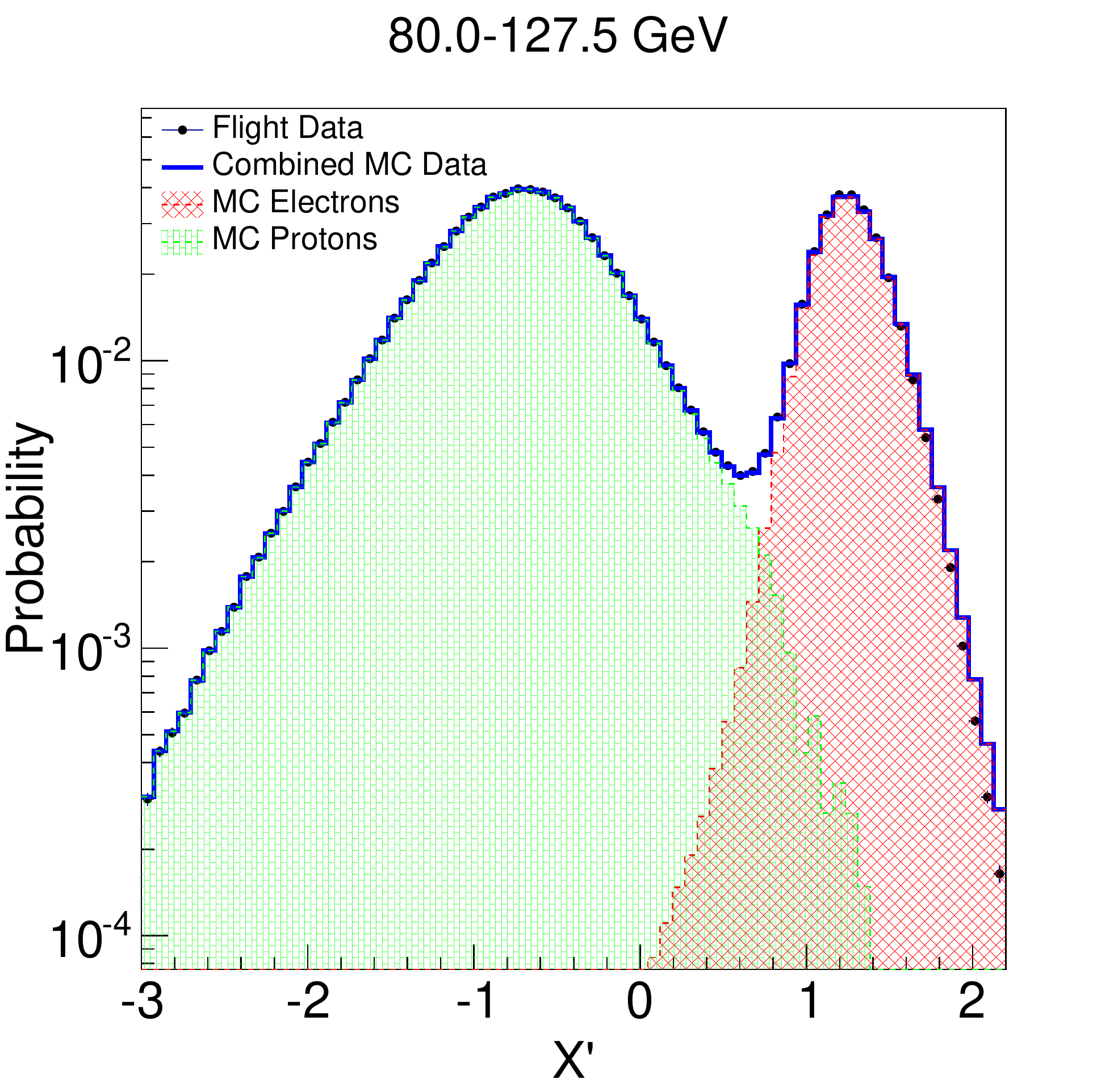}
    \includegraphics[width=0.35\textwidth]{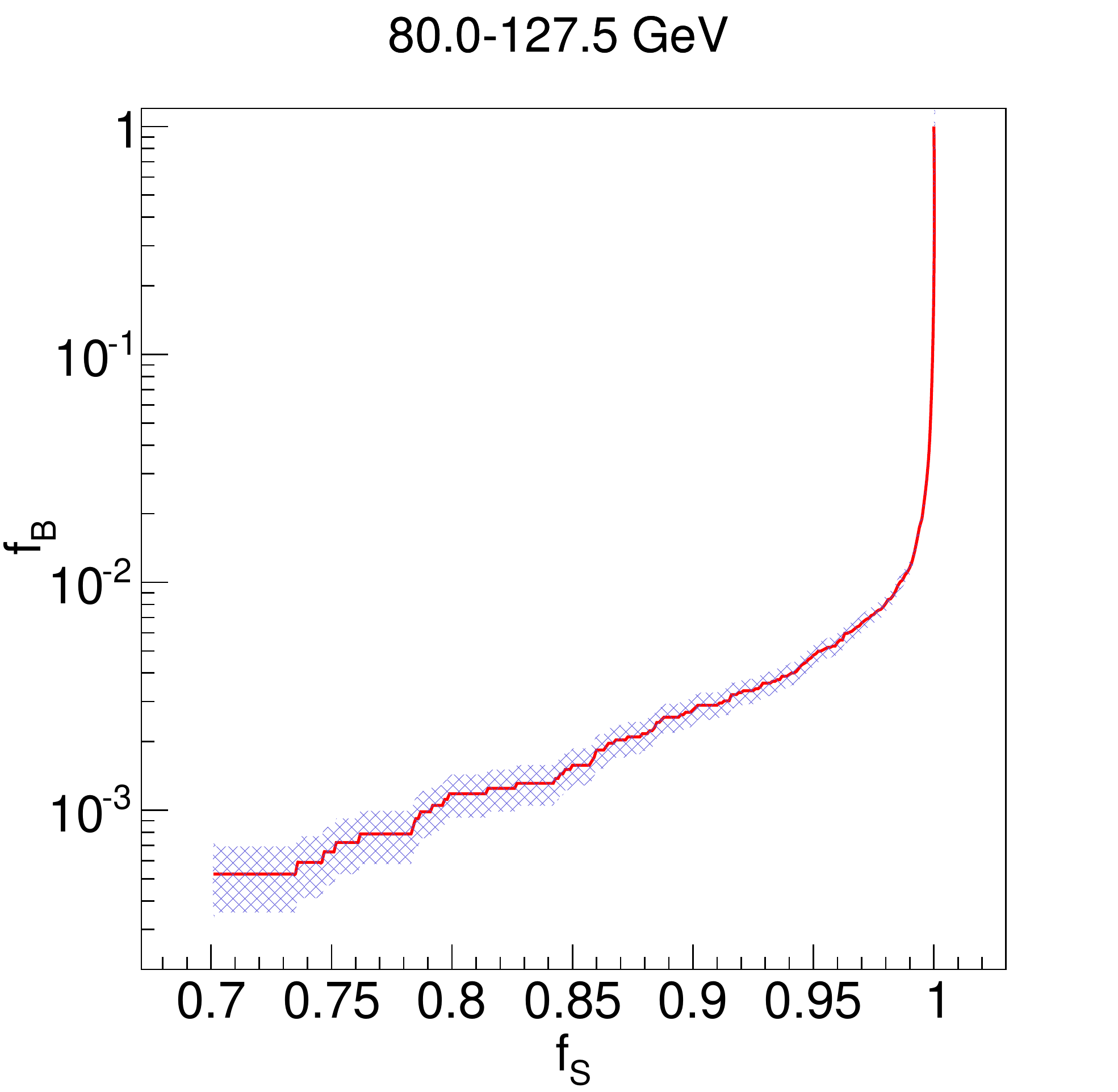}\\
	\includegraphics[width=0.35\textwidth]{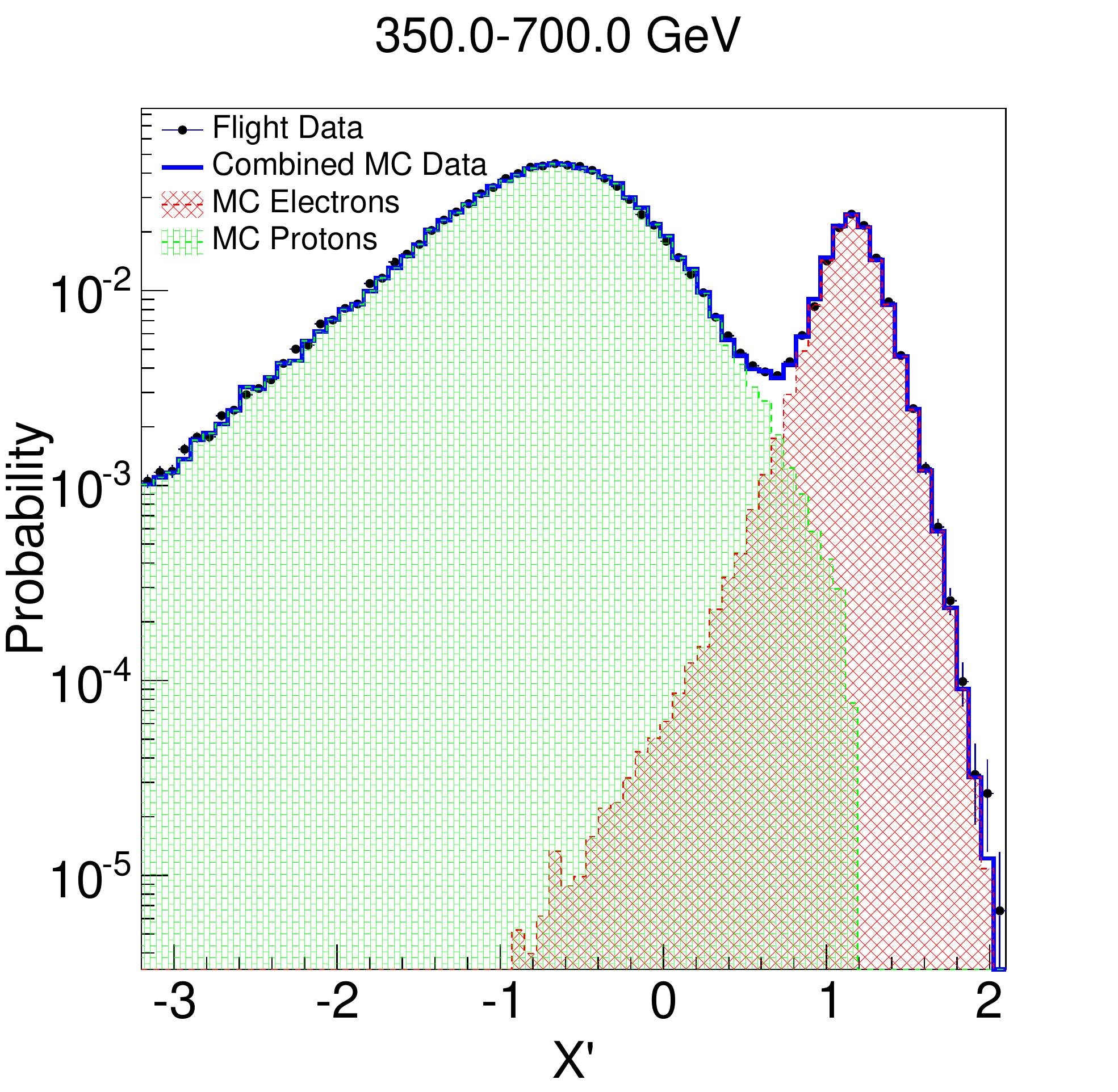}
    \includegraphics[width=0.35\textwidth]{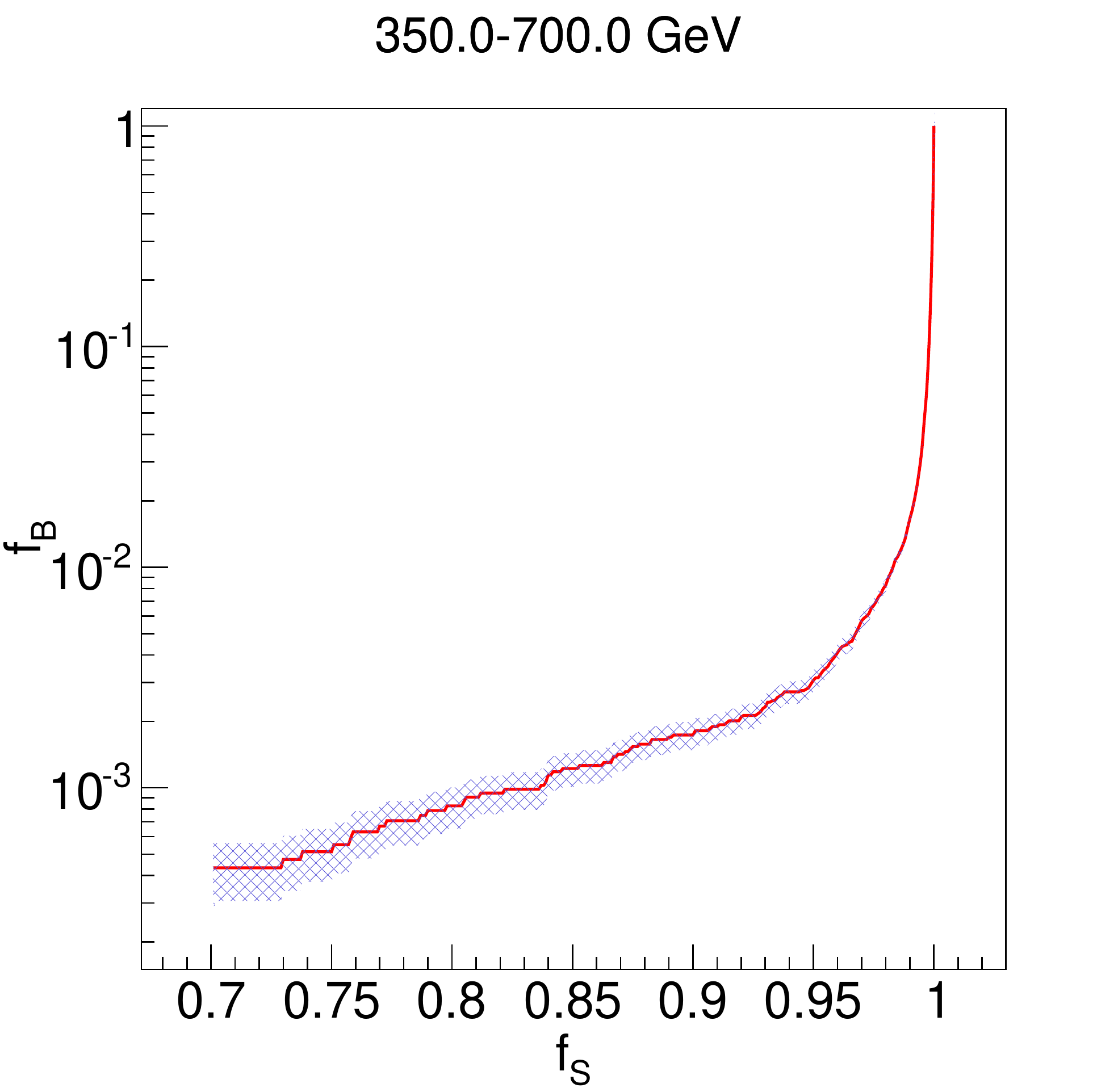}\\
	\includegraphics[width=0.35\textwidth]{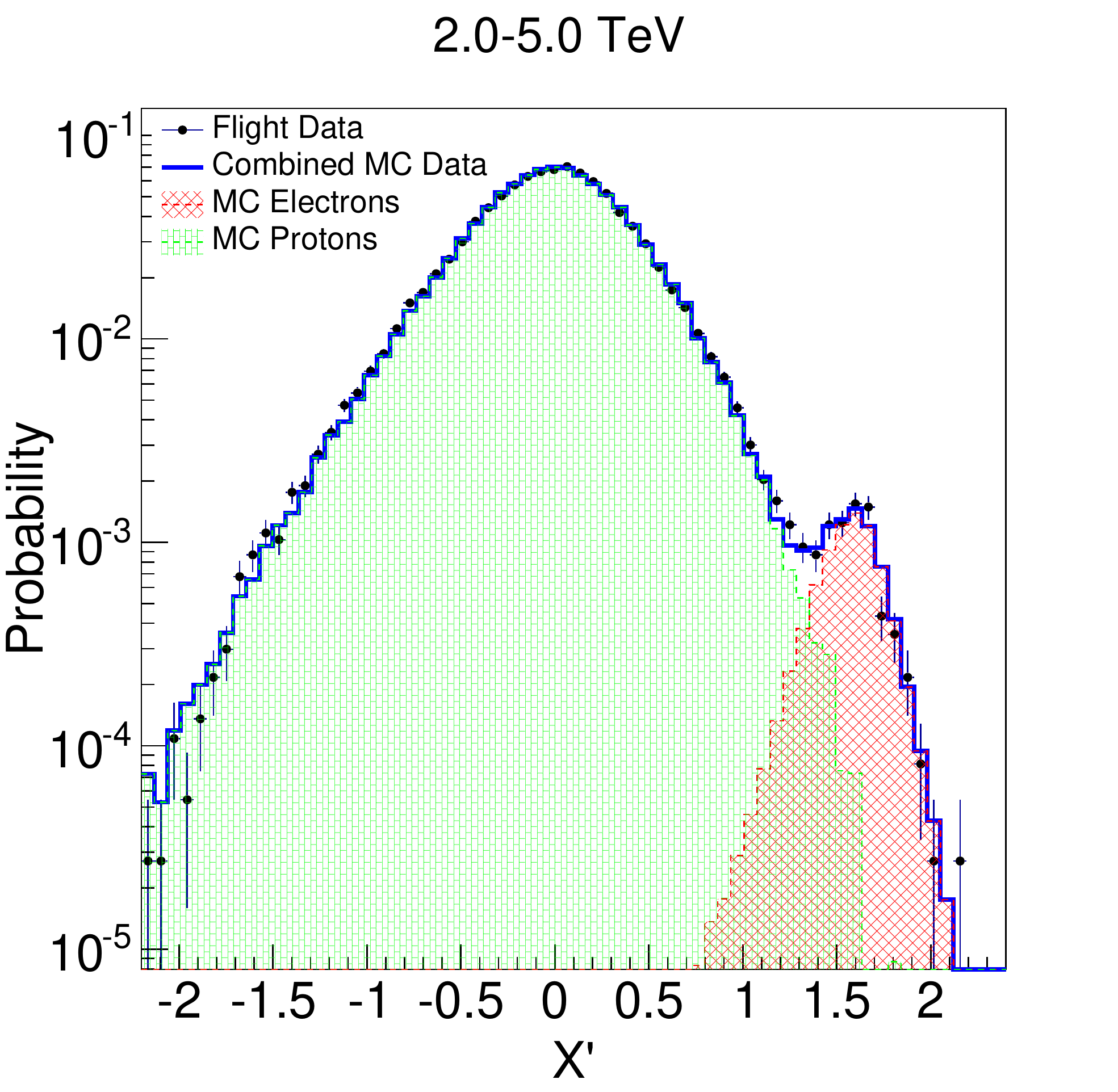}
    \includegraphics[width=0.35\textwidth]{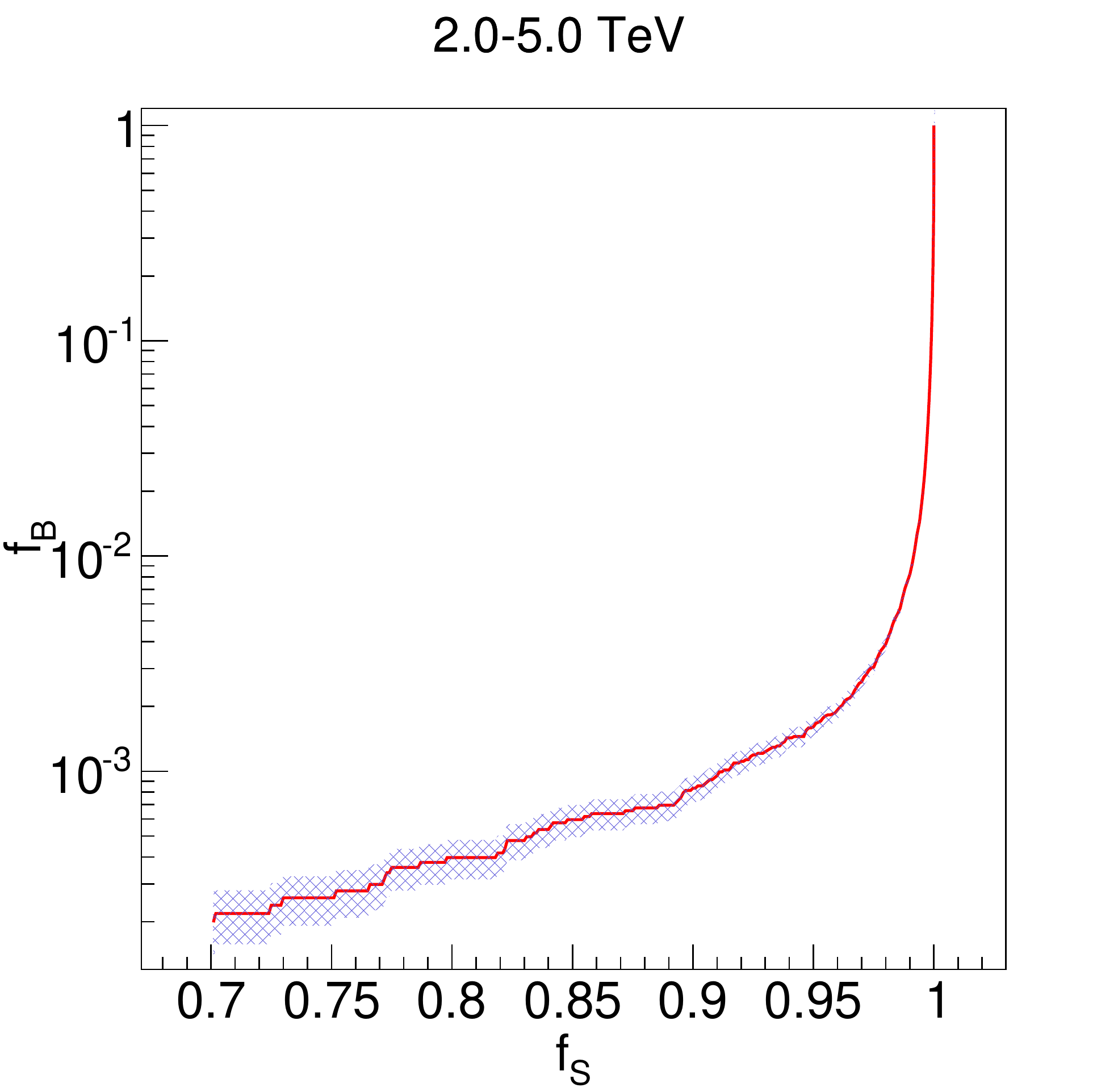}
    \caption{{{Left}:} 
{The distributions} 
 of the rotated first principal component of the flight data
    and fitting results of the MC templates (left panels). {{Right}}: The residual background fractions versus \mbox{signal efficiencies}.}
    \label{fig:Tem_Fit}    
\end{figure}

\textls[-15]{Specifically, we choose three typical reconstructed energy ranges, representing low, middle and high energies, to show the distribution and background estimation.
Comparisons between
the simulation and flight data are shown in the left panels of Figure~\ref{fig:Tem_Fit}
for the three energy bands.
The right panels of Figure~\ref{fig:Tem_Fit} show the relative
efficiencies of protons ($f_B$) and electrons ($f_S$) for different cuts of the $X'$. 
From the template fitting results, we can estimate the residual background fractions given 
signal efficiencies. 
If we set 90\% electron efficiency, the proton contamination is found to be
(0.45 $\pm$ 0.02)$\%$, (0.52 $\pm$ 0.04)$\%$, and (10.55 $\pm$ 1.80)$\%$ for reconstructed energies 80.0--127.5 GeV, 350.0--700.0 GeV, and \mbox{2.0--5.0 TeV,} respectively. }

\begin{figure}[H]
\includegraphics[width=10.5 cm]{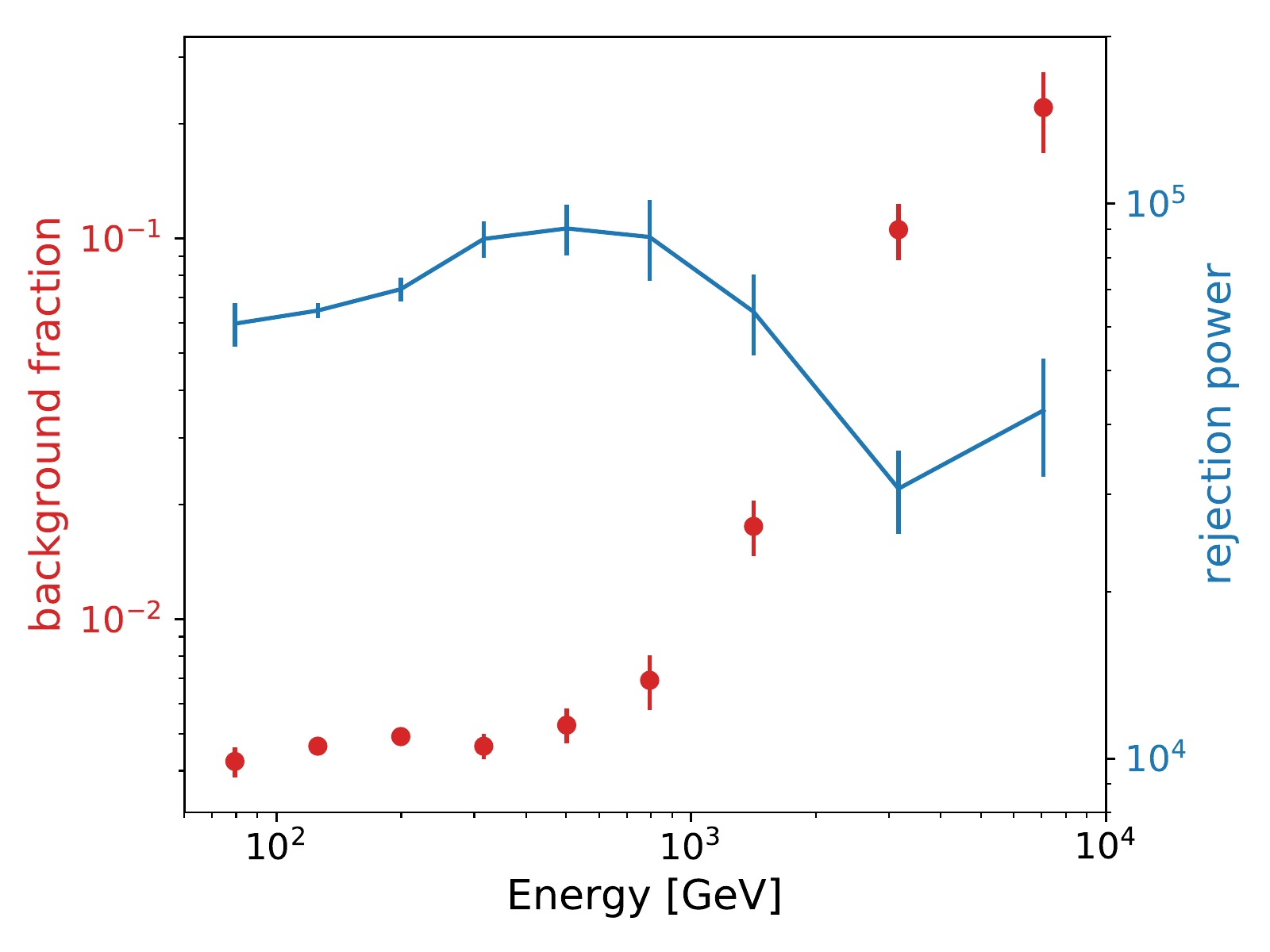}
\caption{
The {background fraction} 
 is shown by red points (left axis) and a rejection power of protons by blue points with a line (right axis).}

\label{fig:Rejection_Power}
\end{figure}

The {background} fraction of protons as
a function of reconstructed event energy is shown in Figure~\ref{fig:Rejection_Power} (left axis). 
{And} for the highest energy range of a few TeVs, {it} is still well controlled 
in our method while keeping a relatively high electron efficiency. As a comparison, the electron 
efficiency decreases significantly above TeV in order to suppress the proton background to a 
level of (10$\sim$20)\% when using the \mbox{traditional method \cite{2017Natur.552...63D}.}

Finally, we obtain the rejection power of protons of the PCA algorithm. The proton rejection power 
is defined as $Q=f_p^{-1}\times\phi_p/\phi_e$, where $f_p$ is the residual proton fraction in 
the electron sample, and $\phi_p$ and $\phi_e$ are the primary fluxes of protons and electrons.
The rejection power is calculated with the reconstructed energy for selected samples and with the primary energy for primary fluxes, respectively. Note that the reconstructed energy corresponds to the primary energy for electrons with a tiny dispersion of $\sim$1\%. For the proton and electron fluxes, we use the fitting results as 
$\phi_p(E)=7.58\times10^{-5}(E/{\rm TeV})^{-2.772}[1+(E/0.48~{\rm TeV})^5]^{0.173/5}$
GeV$^{-1}$~m$^{-2}$~s$^{-1}$~sr$^{-1}$ \cite{2019SciA....5.3793A}, and
$\phi_e(E)=1.62\times10^{-4}(E/{\rm 0.1~TeV})^{-3.09}[1+(E/0.91~{\rm TeV})^{8.3}]^{-0.1}$
GeV$^{-1}$~m$^{-2}$~s$^{-1}$~sr$^{-1}$ \cite{2017Natur.552...63D}.
The proton rejection power as a function of reconstructed event energy is shown in
Figure~\ref{fig:Rejection_Power} (right axis). For the selected three energy bands in 
Figure~\ref{fig:Tem_Fit}, the proton rejection power is $(6.21\pm0.03)\times10^4$, 
$(9.03\pm0.05)\times10^4$, and $(3.06\pm0.32)\times10^4$.

\vspace{-6pt} {}

\section{Conclusions}\label{Sec_Conclusion}
The machine learning methods are more and more widely used in astroparticle physics.
Significant improvements have been achieved in the efficiency and accuracy of particular problems such
as classifications, pattern recognitions, and nonlinear inverting problems. 
Supervised machine learning relies on training, which is based on the simulation data. The advantage 
is that it is not limited by the statistics of the real data, and a very good training of
the model can be achieved. However, this method requires a good match between simulation 
data and real data. As a consequence, the training results are highly model-dependent. 
Unsupervised machine learning, on the other hand, avoids such a model dependence
but is subjected to statistical uncertainties of the experimental data. 

Using an unsupervised machine learning method, the PCA, we discriminate electrons from
protons for the DAMPE experiment. We use the six-year flight data of DAMPE to search for 
effective parameters to distinguish those particles. We find that the PCA method performs
well in the electron identification. The residual proton contamination fraction is estimated
to be $(0.45\pm0.02)\%$, $(0.52\pm0.04)\%$, and $(10.55\pm1.80)\%$ for electron energies of 
80.0--127.5 GeV, 350.0--700.0 GeV, and 2.0--5.0 TeV.
Compared with the traditional method used in Ref.~\cite{2017Natur.552...63D}, the PCA method
improves  the whole energy range. For the same electron efficiency, the proton background
from the PCA method is lower by a factor of two to three. 
Compared with the supervised machine learning method, our approach has a comparable background suppression ability \cite{2021JInst..16P7036D}. 


\vspace{6pt} 



\authorcontributions{Conceptualization, X.L. and M.C.; software, M.C. and Z.X.; investigation, Z.X., X.L., M.C., C.Y., W.J. and W.L.; writing---original draft preparation, Z.X.; writing---review and editing, M.C., X.L., C.Y. and Q.Y.; visualization, Z.X. and Q.Y.; supervision, Q.Y.; project administration, X.L. All authors have read and agreed to the published version of the manuscript.}

\funding{This work is supported by the National Natural Science Foundation of China (Nos. 12173099, 11903084, 12220101003), the Chinese Academy of Sciences (CAS) Project for Young Scientists in Basic Research (No. YSBR-061), the Scientific Instrument Developing Project of the Chinese Academy of Sciences (No. GJJSTD20210009), the Youth Innovation Promotion Association CAS, and the Natural Science Foundation of Jiangsu Province (No. BK20201107).}

\institutionalreview{Not applicable.}

\informedconsent{Not applicable.}

\dataavailability{Not applicable.} 

\acknowledgments{This work uses data recorded by the DAMPE mission, which was funded by the strategic priority science and technology projects in space science of the Chinese Academy of~Sciences.}

\conflictsofinterest{The authors declare no conflict of interest.} 



\abbreviations{Abbreviations}{
The following abbreviations are used in this manuscript:\\

\noindent 
\begin{tabular}{@{}ll}
DAMPE & Dark Matter Particle Explorer\\
CR & Cosmic Rays\\
PSD & Plastic Scintillator Detector\\
STK & Silicon Tungsten tracKer-converter\\
BGO & {${\rm Bi}_4{\rm Ge}_3{\rm O}_{12}$}\\
NUD & Ne{U}tron Detector\\
PCA & Principal Component Analysis\\
MC & Monte Carlo\\
HET & High Energy Trigger\\
RMS & Root Mean Square
\end{tabular}
}

\begin{adjustwidth}{-\extralength}{0cm}
\printendnotes[custom]
\reftitle{References}

\end{adjustwidth}
\end{document}